\documentclass[useAMS,usenatbib,usegraphicx]{mn2e}
\title[Extinction Correction for Type Ia SN Rates. The model]{Extinction Correction for Type Ia Supernova Rates. I.\\ The Model}
\author[M. Riello \& F. Patat]
  {M.~Riello,$^{1,2,3}$\thanks{E-mail: mriello@ast.cam.ac.uk}
  F.~Patat$^1$\\
  $^1$European Southern Observatory,
      Karl-Schwarzschild str. 2, D-85748 Garching bei M\"unchen, Germany\\
  $^2$Dipartimento di Astronomia, Vicolo dell'Osservatorio 2, I-35122 Padova, Italy\\
  $^3$ Institute of Astronomy, Madingley Road, Cambridge CB3 0HA}

\date{Received .................; Accepted ................}

\pagerange{\pageref{firstpage}--\pageref{lastpage}} \pubyear{2004}

\begin{document}

\label{firstpage}

\maketitle

\begin{abstract}
In this paper we present and discuss a new Monte Carlo approach aimed at correcting the observed Supernova (SN) rates for the effects of host galaxy dust extinction. The problem is addressed in a general way and the model includes SN position distributions, SN light curve and spectral library, dust properties and distribution as input ingredients. Even though the recipe we propose is in principle applicable to all SN types, in this paper we illustrate the use of our model only for Type Ia. These represent in fact the simplest test case, basically due to their spectroscopic homogeneity which, to a first approximation, allows one to treat them all in the same way. This test case shows that the final results do not depend critically on the spiral arm dust geometry, while the total amount of dust, its properties and the size of the galactic bulge do have a strong effect. With the availability of more complete spectral libraries and a more accurate knowledge of SN spatial distribution, the method we propose here can be easily extended to Core Collapses events.
\end{abstract}


\section{Introduction}

Until a few years ago, observational measurement of supernova (SN) rates were scanty. The only available measurements were those for local SN rates \citep{stat97,stat99} whilst nothing was known about the evolution with look-back time.
The turning point was represented in the late nineties by the demonstration that Type Ia SNe can be used as precise distance indicators, particularly well suited to probe the geometry of Universe \citep{hzt,riess98,scp}.
The striking result that the Universe is accelerating has indeed 
stimulated the kick off of several new SN search projects, aimed at 
using Type Ia SNe as cosmological probes both to improve the 
determinations of the density parameters $(\Omega_M,\,\Omega_{\Lambda})$ as well as to trace the evolution of the expansion factor through the Universe history \citep[see][for a review]{brunocosmo}. Before that time, studies of SN rates and their evolution with redshift were mainly theoretical speculations.
Today, the few published measurement are based on data from SN searches designed for cosmological purposes \citep{PAIN96,PAIN02,TONRY,goods2}. The only exception is represented by \citet{stat04}, a unique SN search specifically designed to measure the SN rate at intermediate redshifts and with no bias towards Ia's.

A major point of concern when deriving observational estimates of SN rates is the correction for all possible biases affecting the observed number counts. Nowadays, the main source of bias is caused by the effect of dust extinction in the SN host galaxies.
This is indeed severe, particularly for core collapse (CC) SNe, which are preferentially located in dust-rich environments. On the contrary, the effect for Type Ia SNe is usually assumed to be milder, particularly for those exploding in elliptical and early spiral galaxies, which are thought to be almost dust-free environments \citep[see e.g.][]{riess}.
Nevertheless, there are several examples of Type Ia SNe spanning a large range of extinction from moderate to high values: e.g. SN~1994D $A_V=0.186$ mag \citep{1994D}, SN~2002er $A_V=0.62$ mag \citep{julio}, SN~2002bo $A_V=1.364$ mag \citep{2002bo}, SN~2003cg $A_V\simeq2.3$ mag (Elias De La Rosa, private communication) and SN~2002cv $A_V\simeq8.0$ mag \citep{2002cv}.
There are several methods to estimate the amount of extinction suffered by a SN: they require either the availability of the SN light curve (LC) in at least two passbands (around maximum light or between 30 and 90 days past B maximum light) or at least one spectrum with good S/N \citep[e.g. see][for an application]{julio}.
These techniques are nevertheless useless when one needs to estimate the bias introduced by host galaxy extinction on a sample of SNe discovered by a search program. Indeed in such a case, the valuable missing piece of information is a statistical correction factor that would include, in the SN number counts, the contribution of those objects that were extinguished by the dust of their parent galaxy  and consequently weakened below the search detection limit.

So far the only work aimed at studying the effect of extinction on the radial distributions of SN properties in their parent galaxy is represented by the model of \citet*[][hereafter HBD98]{hatano}. Using a Monte Carlo approach based on assumed distributions of dust and SN populations, they derived the expected extinction distributions and projected radial distributions of both Type Ia and CC SNe. 
In particular, they adopted a simple description of the dust distribution from which they derived the total $B$-band extinction per kpc in the plane of the disk, $\alpha(r)$, which is a linear function of $r$ \citep[see Eq. 1 and 2 of][]{hatano}.
In such a parametrisation, the dust density rises from zero in the centre of the model to a maximum value at 5 kpc and then falls to zero at 17.5 kpc.
They found that as the galaxy inclination increases, the projected distribution's peak moves toward the central regions. Moreover, they conclude that the effects of extinction induced by the host galaxy are likely to be the cause of the observed difference in absolute magnitude dispersion of Type Ia SNe within and beyond 7.5 kpc, confirming the findings of \citet{wang}.

Despite the obvious interest in the observed spatial dependence of SNe properties, in this work we will rather focus on deriving a recipe to correct the observed SN rate for host galaxy extinction. The HBD98 model has been used to derive extinction corrections for both local \citep{stat99} and high-$z$ \citep{TONRY} SN rates. In particular, \citet{stat99} have carefully tested the HBD98 approach to derive the extinction corrections as a function of galaxy inclination. Interestingly enough, they found that the model appears to overcorrect the SN rate of distant galaxies suggesting that a more detailed modelling of the dust distribution and properties is possibly required.
All modern measurement of SN rates at high-$z$ currently available have either not been corrected for dust extinction \citep{stat04,PAIN96,PAIN02,TONRY} or were corrected \citep{goods2} using the HBD98 approach. 

For these reasons, and given the importance of this topic in modern astrophysics, we have decided to address the problem with a more detailed approach.
Our final goal is to construct a reliable model which is able to provide the "bias factor" needed to correct the observed SN rates for dust extinction. In this framework, we started from the Monte Carlo approach of HBD98, which clearly has proven to be effective, and tried to construct a more general model able to cope with the variety of dust properties and distributions that are observed in different spiral galaxies. This is indeed a key feature to enable a reliable determination of the dust bias factor for a real SN search, where the galaxy sample is clearly a mixture of types and morphologies. 
Recently \citet{commins} developed some models for extinction of Type Ia SNe due to dust in spiral galaxies. The main purpose of his work is attempting to account for observational effects that play a significant role in searches for high redshift Type Ia SNe. In particular, that paper investigates the expected dependence of the observed amount of extinction in high redshift Type Ia SNe rather than trying to quantify the effect of dust extinction on the measured SN rates. In this respect, we feel that the study of \citet{commins} and the present work are complementary. Nevertheless we will discuss the main differences between our model and Commins's one in Section \ref{par:method}.

In this paper we present the model and the general recipe to derive the extinction correction factor while the application to the computation of extinction-corrected SN rates will be discussed in a companion paper. In Section \ref{par:method} we outline the model and the assumed spatial distributions of dust and SNe. Section \ref{par:parspace}  presents model predictions of the extinction distributions and projected radial distributions together with their dependence on model parameters. As an illustration we show the effects of dust for a single test case in Section \ref{par:app}. Section \ref{par:discussion} presents a discussion of the physical parameters which have a major impact on the extinction correction factor.
Finally we summarise our results in Section \ref{par:end}.

\section{Method}\label{par:method}

The method we follow in this work is based on the Monte Carlo
technique and it consists of a few basic steps grouped in two main stages: the Monte Carlo and the Integration part. In the first stage Type Ia SN positions inside host galaxies are generated according to different distribution functions (see Sec. \ref{par:sndists}). After that, random SN phases and lines of sight are uniformly and isotropically generated, respectively. These steps complete the Monte Carlo stage of the simulation. The Integration stage uses the Monte Carlo data to compute the SN magnitudes. First the total column density along the line of sight associated to each object is computed by integrating the dust density function (see Sec. \ref{dust}) along the specific line-of-sight. Finally, the apparent magnitudes of SNe are computed integrating template SN spectra, redshifted to the suitable redshift $z$, over the selected passband $\mathcal{P}$ in the presence, or absence, of dust absorption using Draine's extinction laws \citep{draine} (see Sec. \ref{snmag}). 
In this way we can study how the distribution of SN extinction depends on different geometries (e.g. galaxy inclination, different SN distributions inside the host galaxy, etc.) and different physical parameters (e.g. face-on central optical depth, etc.). In this work we have always used Draine's extinction law for $R_V=3.1$ but the model implementation considers also the cases $R_V=4.0$ and $R_V=5.5$. In principle, any another extinction law for which the extinction coefficients as a function of wavelength are provided is suitable as well. We note that, from the analysis of individual SNe, in some cases values of $R_V$ have been found as low as 2.0 (Elias de la Rosa, private communication). Though the approach we used is general, we decided to limit our discussion only to Type Ia SNe both because of the lack of enough data (LC and spectra) for each CC SNe type (Ib, Ic, IIP, IIL, etc.) and because of the present great interest in Type Ia SNe due to their cosmological applications.

Throughout this paper we will assume the standard cosmological model ($H_0=71$ km s$^{-1}$ Mpc$^{-1}$, $\Omega_M=0.3$, $\Omega_\Lambda=0.7$). The adopted values for the other parameters will be discussed in the following sections.

\subsection{Type Ia SNe distribution}\label{par:sndists}

It is generally accepted that Type Ia SNe originate from the
thermonuclear explosion of accreting white dwarfs (WD) in binary
systems. The debate on the nature of the progenitor systems is still
active. In particular there are two main scenarios: (1) the {\em
  single degenerate}, in which the progenitor WD accretes
hydrogen-rich matter via mass transfer from a companion main sequence
or low mass RGB star \citep{whelaniben, ibentutukov}; (2) the {\em
  double degenerate}, in which the binary system is composed of two
WDs, the exploding WD reaches the Chandrasekhar limit and explosive
carbon ignition occurs at its centre \citep{ibentutukov}.

It thus seems reasonable to assume that the old galaxy stellar component is a good tracer of the Type Ia SNe spatial distribution.
The stellar component in a spiral galaxy can be described by two structures: a spheroidal bulge and a disk. To model these two components, we mainly followed the method employed in radiative transfer codes used to derive extinction and polarization effects in dusty spiral galaxies, (see e.g. \citealt*{bianchi}, \citealt{xilouris}). We hence used the luminosity density distributions for the two galaxy components to describe the spatial number density of Type Ia SNe, which we will denote as $\rho_{\rm Ia}$ \citep[see also][]{commins}.

To reproduce the variety of spiral Hubble types, we have considered different bulge-to-total luminosity ratios $(B/T)$ in the range 0 (resembling an Sd galaxy) to 0.5 (resembling an Sa). Given the parallelism between the luminosity distribution and the Type Ia SN spatial distribution previously introduced, once the total number, $N_{\rm Ia}^{\rm tot}$ of simulated Type Ia SNe has been fixed, a fraction $B/T$ is generated according to the bulge distribution and the remaining fraction $(1-B/T)$ is generated according to the disk distribution.

We note that the properties of the assumed spatial distribution of Type Ia SNe are among the most important parameters of our model. Although the adopted spatial distribution is rather simple, it takes into account the diffent scale height of dust and SNe suggested by the observations \citep{DVP92,LE}. A more detailed treatment of the differences between the distribution of the stellar component and Type Ia SNe could improve the model but requires a careful analysis of the differences in the distributions between single stars and primordial binaries which is beyond the scope of the present study.

\subsubsection{Bulge distribution}\label{bulge}

For the bulge, following \citet{bianchi}, we adopt a Jaffe distribution \citep{jaffe}:
\begin{equation}
\rho_{\rm Ia}^b(r)=\frac{\rho_0^b}
{\left(r/r_{\rm Ia}^b\right)^2
\left[1+\left(r/r_{\rm Ia}^b\right)\right]^2}
\end{equation}
where $r$ is the physical distance from the galactic centre and
$r_{\rm Ia}^b$ is the scale radius. The volume integral of the Jaffe
distribution converges for $r\to\infty$ but clearly, for numerical
reasons, we need to introduce a truncation radius $r_{\rm
  max}^{b}=nr_{\rm Ia}^b$. Without this truncation, at large radii the
number of bulge SNe would indeed dominate once again that of the disk. In this work we used $r_{\rm Ia}^b=1.87$ for the bulge scale-length and $n=8$ for the truncation radius. This value of $r_{\rm Ia}^b$, which corresponds to an equivalent radius $R_e=1.60$, is close to the one inferred for our Galaxy \citep{bianchi}.

\subsubsection{Exponential disk distribution}\label{expdisk}

As for the bulge, we assume that the spatial distribution of stars in the disk is traced by the luminosity density. As a first approximation we can consider a double exponential law \citep{freeman}, namely
\begin{equation}
\rho_{\rm Ia}^d(R,z) = \rho_0^d\exp{\left(-\frac{R}{R_{\rm Ia}^d}-
\frac{|z|}{z_{\rm Ia}^d}\right)}
\label{eq:plaindisk}
\end{equation}
where $R$ and $z$ are the cylindrical coordinates, $R_{\rm Ia}^d$ is
the disk scale-length and $z_d$ the disk scale-height. Following the
same approach as in the bulge case, we introduce a radial, $R_{\rm
  max}^{{\rm Ia},d}=nR_{\rm Ia}^d$, and a vertical, $z_{\rm max}^{{\rm Ia},d}=mz_d$, external truncation. In this work we used $r_{\rm Ia}^d=4.0$ kpc and $z_{\rm Ia}^d=0.35$ kpc which are similar to the values inferred for the old disk population of the Galaxy. For the outer edge truncation we adopted $n=m=6$ \citep{bianchi}.

\subsubsection{Perturbed exponential disk distribution}
\label{spiraldiskIa}

As the aim of this work is to improve the calculation of the statistical properties of Type Ia SNe extinction by host galaxy dust, it is crucial to simulate their spatial distribution inside the galaxy as realistically as possible.
In particular, in their study, \cite{mcmillan} considered a sample of 35
spectroscopically identified Type Ia SNe and measured their positions
relative to the spiral arms from a homogeneous data set of $V$ and $I$ images. Interestingly enough, they found that Type Ia SNe occur closer to the arms than a perfectly randomised disk population. However, they note that Type Ia SNe are not statistically distinguishable from the distribution of stellar light around the arms (which should be considered as a tracer of the stellar density). 

We decided thus to implement in our model both cases: (1) no 
concentration closer to arms, which we achieve by distributing the 
SNe in a homogeneous exponential disk (Eq. \ref{eq:plaindisk}); (2) concentration closer to arms, which we achieve by simulating a spiral structure as a perturbation of a homogeneous exponential disk. 
A realistic spiral structure can be represented by logarithmic spiral arms \citep{bm} as a perturbation on a plain exponential disk. Following
\cite{misiriotis}, for the perturbed distribution we adopt the
following expression:
\begin{equation}
\tilde{\rho}_{\rm Ia}^d(R, z,\phi)=
\rho_{\rm Ia}^d(R,z)\left[
1+w\sin{\left(\frac{N_a}{\tan{p}}\log{R}-N_a\phi\right)}
\right]
\label{eq:spiraldisk}
\end{equation}
where $R,z$ and $\phi$ are the cylindrical coordinates. $\tilde{\rho}_{\rm Ia}^d(R,z)$ is made up of two components: the plain exponential disk $\rho_{\rm Ia}^d$, given by Eq. \ref{eq:plaindisk}, and the spiral perturbation represented by the term in square brackets. The three parameters governing the spiral perturbation are: $w$, the spiral perturbation amplitude parameter ($w=0$ correspond to the plain exponential disk); $p$, the pitch angle (low $p$ values produce tightly wound arms) and $N_a$, the number of spiral arms.

Though we have implemented in our model the perturbed spiral distribution described by Eq. \ref{eq:spiraldisk}, all the results discussed in this work are based on simulations in which the disk component was described by Eq. \ref{eq:plaindisk}. We believe indeed that, though the evidences presented by \citet{mcmillan} appear quite convincing, observational data is still too scanty to permit modelling without "blind" assumptions. We thus prefer to limit the study to the simpler case to reduce the number of low-impact model's free parameters.

\subsection{Dust distribution}\label{dust}

\begin{figure*}
%
% The following figure was produced with:
% analsim.x -tauplot -tau 1.0 -draine 0 -binsize 0.1 -narms 2 -ws 0.4
%		-pitch 10. -bw -distype eps -plotfile S21__t1__hr__bw
%
\includegraphics[width=84mm]{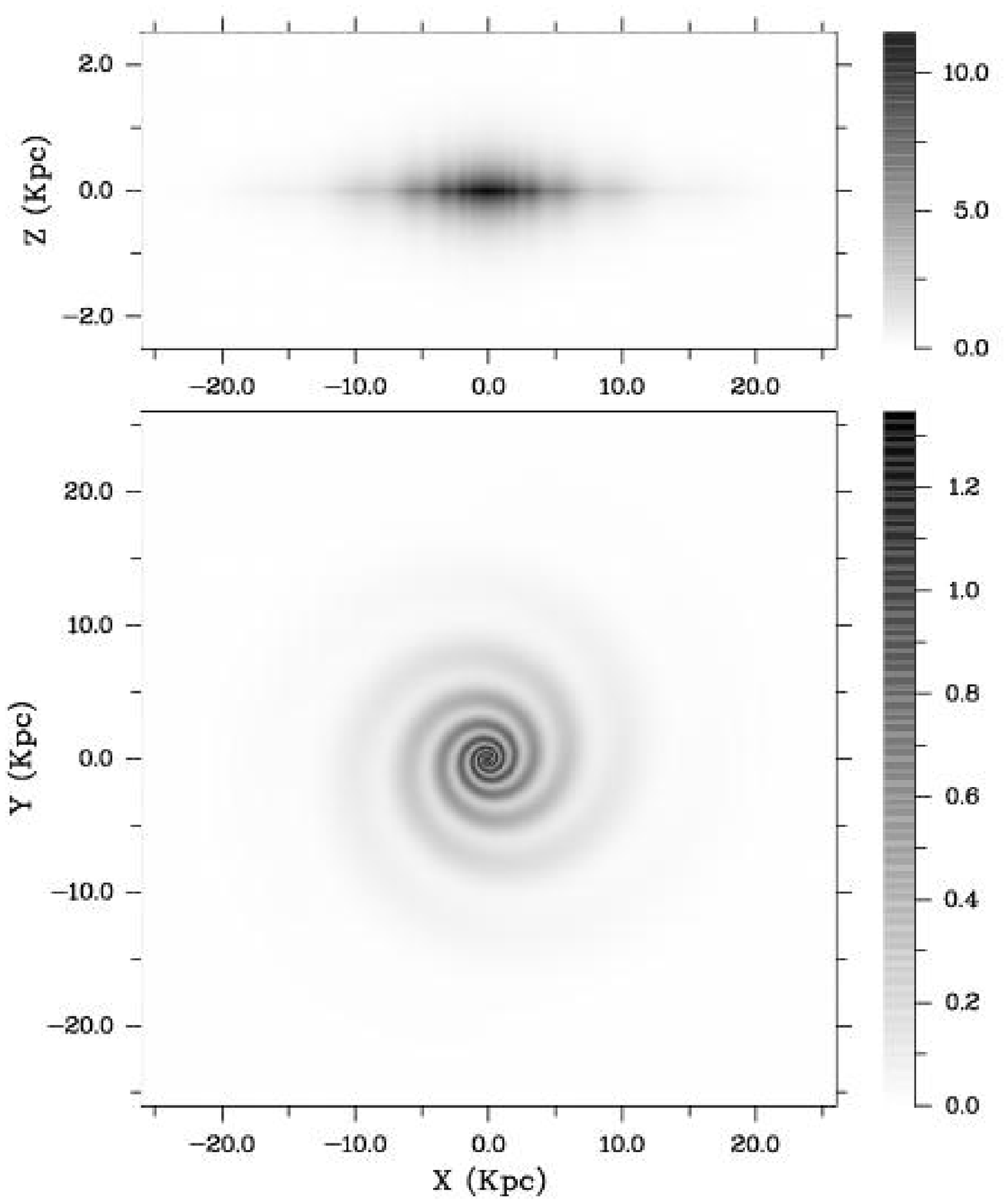}
\includegraphics[width=84mm]{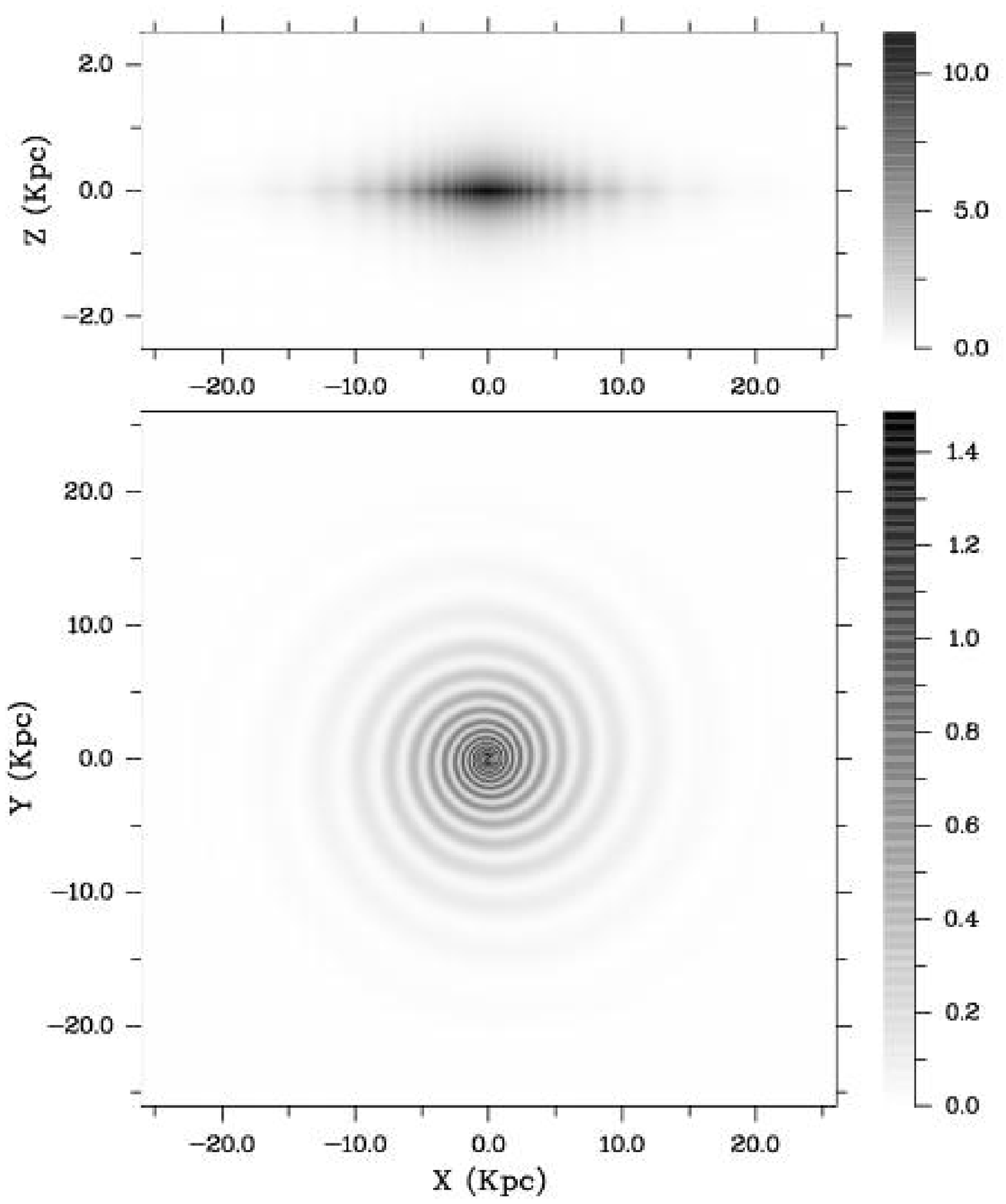}\\
\includegraphics[width=84mm]{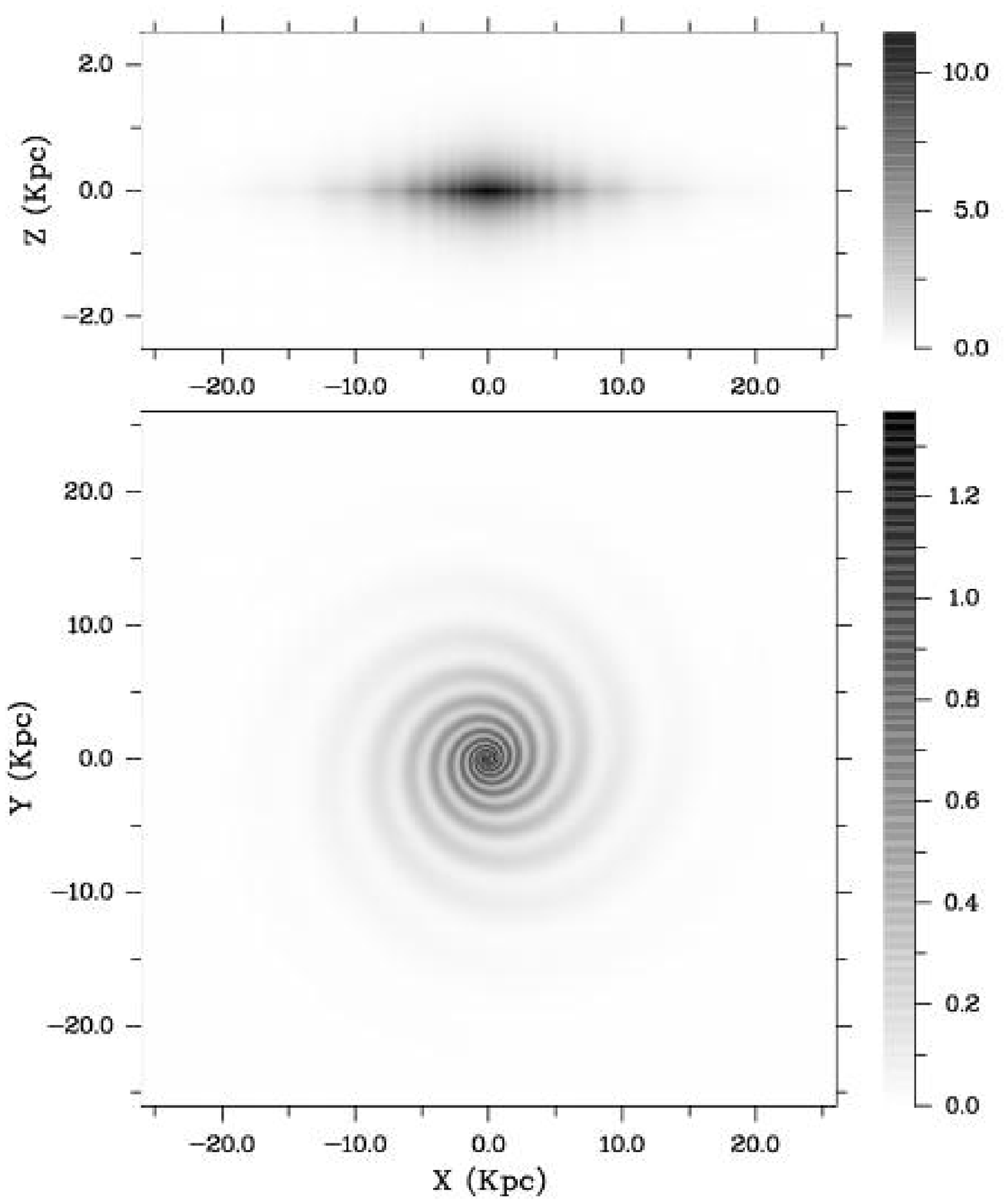}
\includegraphics[width=84mm]{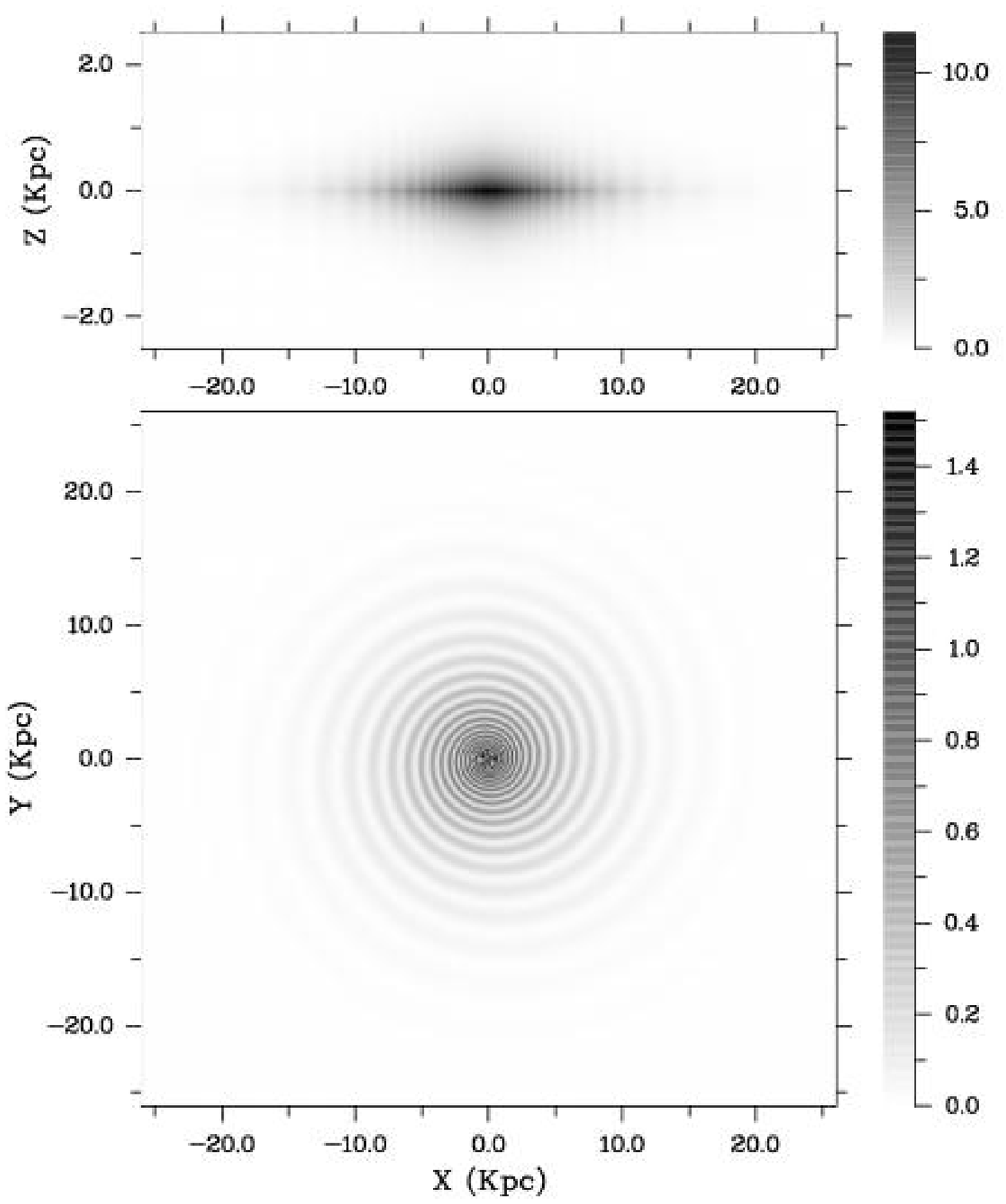}
\caption{Dust distribution for the four spiral models considered in
  this work from left to right and top to bottom the total optical
  depth map is shown for models S21, S22, S31, S32 with $\tau_V(0)=1.0$ and $B/T=0.3$. See the text and Table \ref{dustpars} for details.}
\end{figure*}

The exponential disk is commonly used in radiative transfer
Monte Carlo codes of galaxies because it is easily implemented
numerically and, moreover, it makes simulations very efficient in terms of computation time. In our approach we do not solve for the radiation transfer and thus we can implement more realistic dust distributions including a spiral arm structure, still keeping an affordable computation time.
The dust density distribution $\tilde{\rho}_D(R,z,\phi)$ is described by an expression analogous to Eq. \ref{eq:spiraldisk}, where the choice of the spiral perturbation parameters ($N_a$, $p$ and $w$) and the plain disk scale length and height ($R_D^d$ and $z_D^d$, analogous to $R_{\rm Ia}^d$ and $z_{\rm Ia}^d$ in Eq. \ref{eq:plaindisk}) are obviously independent from that used for Type Ia SNe distribution.
In analogy with the Type Ia SNe case, we have introduced a radial, $R_{\rm max}^D=nR_D$, and a vertical, $z_{\rm max}^D=mz_D$, external truncation. In this work we adopt $R_D=4.0$ kpc and $z_D=0.14$ kpc and the same truncation factors as for the Type Ia SN distribution (see Sec. \ref{expdisk}).
The total content of dust in a galaxy is then normalised in terms of the optical depth along the galaxy symmetry axis:
\begin{equation}
\tau_{\mathcal{P}}(0)=\int_{-\infty}^{+\infty}{\rho_D^d(R=0, z) dz}=
\rho^d_0\left[2z^de^{-m}C_{ext}(\lambda_{\mathcal{P}}^{\rm eff})\right]
\label{eq:dustnorm}
\end{equation}
where $m$ is the disk vertical external truncation introduced before and the subscript $\mathcal{P}$ indicates that the normalisation depends on the adopted passband, which is characterised by the value of the extinction coefficient at its effective wavelength $\lambda_{\mathcal{P}}^{\rm eff}$. Once $\tau_{\mathcal{P}}(0)$ has been fixed, Eq. \ref{eq:dustnorm} provides the normalisation $\rho_0^d$ for the dust distribution. 

Eq. \ref{eq:dustnorm} is rigourously valid for the plain disk only, but it has been used also for the perturbed exponential disk because, in such case, the normalisation has not an analytical expression.

We note here that HBD98 did not allow the dust density to increase to the very centre of the galaxy. Conversely our models do not adopt any internal cut-off in the dust distribution. It appears that the central hole is a well know feature of the Milky Way dust distribution \citep{milkyway}. We note however that usually such a central cut-off is not adopted by modern radiative transfer models used to fit the optical and near infrared surface brightness of observed galaxies. Therefore, for the lack of compelling evidence to the contrary and for the sake of generality, we assume a smooth dust distribution without a central cut-off as described above.
This difference may partially justify some of the diversities we found between our results and HBD98's ones (see Section \ref{par:extinc}).

\subsection{Type Ia SN apparent magnitudes}\label{snmag}

To compute the SN apparent magnitudes one can follow two different
approaches, one based on SN light curve and the other based on SN
spectra.
In the first approach one has to select a particular SN light curve
template, thus choosing a particular $\Delta m_{15}$ parameter (defined as the $B$ band magnitude difference between maximum light and 15 days after). Other parameters are the absolute magnitude at maximum, its intrinsic dispersion and the $K$-correction which depends on redshift, filter and SN type and phase. Finally one has to derive the total extinction along the line of sight and in the selected passband.

In the second approach, which we choose to follow, one has to select
a set of template spectra, covering a reasonable phase range. Then, for a given SN phase, the spectrum is extinguished according to the total optical depth along the line of sight, and integrated over the passband $\mathcal{P}$ to derive the apparent flux $F_{\mathcal{P}}$:
\begin{equation}
F_{\mathcal{P}} =\frac
{\displaystyle
\int_{0}^{+\infty}
{F(\lambda)\exp{\left[-N_dC_{\rm ext}^{H}(\lambda)\right]}
\mathcal{T}_{\mathcal{P}}(\lambda)\,d\lambda}
}
{\displaystyle
\int_{0}^{+\infty}
{\mathcal{T}_{\mathcal{P}}(\lambda)\,d\lambda}
}
\label{eq:specmag}
\end{equation}
where $F(\lambda)$ is the (input) observed SN spectrum, $C_{\rm ext}^{H}(\lambda)$ is the extinction cross section per H nucleon $({\rm cm}^2/{\rm H})$ as a function of wavelength for a given value of $R_V$ (see Sec. \ref{drainecext}), $\mathcal{T}_{\mathcal{P}}$ is the total transmission for the passband $\mathcal{P}$ and $N_d$ is the column density obtained by computing the integral of the dust density function along the line of sight.
The apparent magnitude of the SN is finally computed converting the
flux to magnitudes and applying the photometric zero
point $zp_{\mathcal{P}}$, namely:
\begin{equation}
m_{\mathcal{P}} = -2.5\left\{
\log{\left[F_{\mathcal{P}}(1+z)\right]}
-\log{\left(zp_\mathcal{P}\right)}
\right\}
\label{eq:snmag}
\end{equation}
In this work, we considered only the $V$ passband
\citep{bessell} with a zero point $zp_V=3.55\times10^{-9}$
erg~s$^{-1}$~cm$^{-2}$~\AA$^{-1}$ but the approach is absolutely general.

To compute the SN apparent magnitude from Eq. \ref{eq:snmag} one needs the SN spectrum at a random phase $\phi$ to compute the SN flux through Eq. \ref{eq:specmag}. The problem is clearly that, even in the best cases, SN spectral sequences have a minimum separation of 1 day (and even larger, particularly at later phases). The approach we adopted is to create a look-up table which enable one to construct an interpolated spectrum at any phase in the range spanned by the original input spectra, so that dangerous extrapolations are avoided. This procedure is described in the next section.

\subsection{Spectral Library}\label{par:input}

The SN spectral library we used is composed by 28 spectra of SN 1998aq \citep{1998aq}, merged with some data from SN 1994D \citep{1994D} and SN 1992A \citep{1992A1,1992A2} spanning the phase range $-9\leq t\leq90$ days from maximum light. This was possible because these three objects are all "normal", very similar, Type Ia SNe.
With the aim of increasing the generality of our approach, we devised a recipe that enables the use of a composite spectral library, i.e. a collection of spectra from different objects. One important fact to keep in mind is that dust extinction is a local process, i.e. it is working at the SN rest frame.
All input spectra are first corrected to redshift z=0 (i.e. the SN rest frame), trimmed to the overlapping wavelength range and re-binned to a common wavelength step $\Delta\lambda$.
The last two operations have the only purpose of making the model implementation easier. The next step is to ensure the consistency of the spectrophotometry with the LC: for each spectrum (i.e. phase) a flux normalisation is computed in the selected passband $\mathcal{P}$, which ensures that the magnitude obtained from the spectrum (through Eqs. \ref{eq:specmag} and \ref{eq:snmag}) is equal to the one derived from the SN LC template.

When computing the apparent flux $F_{\mathcal{P}}$ using Eq. \ref{eq:specmag}, the spectrum is first extinguished and then redshifted to $z_{\rm sim}$, the redshift at which the simulation is focused.
We note that, within this approach, the usable phase range is given by the intersection of the light curve template and the spectral library ones. In particular, in this work the simulation phase range (in the observer's rest frame) is: $-5\leq\phi\leq80$.

\subsection[]{The extinction coefficient $C_{\rm ext}^H(\lambda)$}\label{drainecext}

\cite{wd} and \cite{ld} have developed a carbonaceous-silicate grain
model which successfully reproduces observed interstellar extinction,
scattering, and infrared emission for the Milky Way. In his review, \cite{draine},
discussed the available evidence for dust extinction per unit H column density in regions with different extinction curves. Based on this discussion, he has renormalized the size distributions of \cite{wd}.

In this work we used the extinction cross-sections per unit H nucleon $C_{\rm ext}^H$ for $R_V=3.1$, which is considered to be appropriate for the typical diffuse H{\sc i} clouds in the Milky Way. The model can be used as well with the extinction cross-sections computed for $R_V=4.0$ and $R_V=5.5$.

Further details on the grain abundance per H nucleon adopted for the three different $R_V$ values and the models themselves can be found at Draine's web page\footnote{http://www.princeton.edu/$\sim$draine}.

\section{Exploring the parameters' space}\label{par:parspace}

\begin{table}
\caption{Model parameters for the Type Ia SN and dust distribution
  used in this work.}
\label{modpars}
\begin{tabular}{@{}lcccc}
\hline
Component & $r_0$ & $r_{\rm max}$ & $z_0$ & $z_{\rm max}$\\
\hline
Ia Bulge & 1.87 kpc & $8r_0$ & - & -\\
Ia Disk & 4.00 kpc & $6r_0$ & 0.35 kpc & $6z_0$\\
Dust Disk & 4.00 kpc & $6r_0$ & 0.14 kpc & $6z_0$\\
\hline
\end{tabular}
\end{table}

The model outlined in the previous section can reproduce several physical situations by means of different combinations of physical
and geometrical parameters. In particular:
\begin{enumerate}
\item
{\em Dust spiral arms geometry}. The geometry is governed by the
perturbation parameter $w_d$, the number of arms $N_a$ and the pitch angle $p$. We considered five combinations to describe a plain exponential disk, a normal and an extreme spiral with two or three arms. The parameters used in each model and the corresponding IDs are summarized in Table \ref{dustpars}.

\item
{\em Total optical depth along the symmetry axis} $\tau_V(0)$. 
This parameter is used to fix the total amount of dust content of the galaxy in which SNe are being simulated. All the simulations in this work are for $\tau_V(0)=1.0$ because it is then easy to re-scale the results to different normalisation values. Moreover this is also considered a typical value \citep{bianchirev}.

\item
{\em Bulge-to-Total ratio}. Our assumption that the distribution of Type Ia SNe follows that of stars, implies that a fraction $B/T$ of the SNe will be distributed in the bulge, while the remaining $(1-B/T)$ will be distributed in the disk. We considered three different cases for $B/T$: 0.0, 0.3, 0.5. 
The first one describes a system in which all the simulated Type Ia SNe are located in the disk while the last one represents a system in which half of the simulated Type Ia SNe are located in the bulge and half in the disk.

\item
$R_V$. The value assumed for $R_V$ depends on the interstellar environment intersected by the line of sight: extinction curves which are measured in low density interstellar environments are usually characterised by small values of $R_V$ (e.g. 3.1) whilst the value assumed by $R_V$ is higher ($4<R_V<6$) if the line of sight is intersecting denser clouds. The mean extinction curve commonly used has $R_V=3.1$; we thus adopted this value in all our simulations.
\end{enumerate}

\begin{table}
\caption{Dust geometries.}
\label{dustpars}
\begin{tabular}{@{}lcccc}
\hline
Model & $N_a$ & $w_d$ & $p$ & Note\\
\hline
S0 & 0 & - & - & plain disk\\
S21 & 2 & 0.4 & 10$^\circ$ & normal spiral\\
S22 & 2 & 0.6 & 5$^\circ$ & extreme spiral\\
S31 & 3 & 0.4 & 10$^\circ$ & normal spiral\\
S32 & 3 & 0.6 & 5$^\circ$ & extreme spiral\\
\hline
\end{tabular}
\end{table}

\subsection{Extinction distribution as a function  of galaxy inclination}
\label{par:extinc}

\begin{figure}
%
% The following figure was produced with:
% analsim.x -extdist S21-3.1__t1.0__b0.3.sim - avmax 9.0 -paper \
%		-distype eps -plotfile S21-3.1__t1.0__b0.3__avdist
%
\includegraphics[width=84mm]{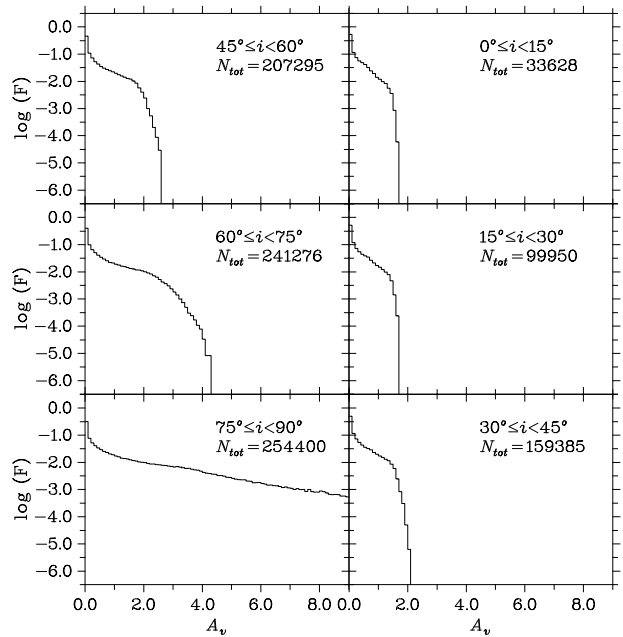}
\caption{Distribution of the total extinction $A_V$ for different galaxy inclination angles for a S21 model (see Tab. \ref{dustpars}) with $\tau_V(0)=1.0$ and $B/T=0.3$. The fraction $F$ of objects in the inclination bin considered is plotted in log scale. In each panel the range of inclinations considered and the total number of simulated SNe is shown.}
\label{avinc}
\end{figure}

\begin{table*}
%
% The data in this table was generated with:
%	analsim.x S21-3.1__t1.0__b0.0__z0.10__1E6.sim -hstats
%	analsim.x S21-3.1__t1.0__b0.3__z0.10__1E6.sim -hstats
%	analsim.x S21-3.1__t1.0__b1.0__z0.10__1E6.sim -hstats
%
\caption{Extinctions for different inclination bins. $\langle A\rangle$ indicate the mean extinction value in the $V$ band. The subscripts $f$ and $l$ indicate that the mean refers to the whole sample and to an extinction-limited subset with $A<0.6$ respectively. The table shows the figures for three different SN populations: $B/T=0.0$ (disk), $B/T=0.3$ and $B/T=1.0$ (bulge). The maximum extinction value $A_{\rm max}$ and the fraction of SNe, $F$, in the extinction-limited subsets are also shown.}\label{hatanostats}
\begin{tabular}{lcccccccccccc}
\hline
$B/T\longrightarrow$ & 0.0 & 0.3 & 1.0 & 0.0 & 0.3 & 1.0 & 0.0 & 0.3 & 1.0 & 0.0 & 0.3 & 1.0 \\
Inclinations & $\langle A\rangle_f$ & $\langle A\rangle_f$ & $\langle A\rangle_f$ & $\langle A\rangle_l$ & $\langle A\rangle_l$ & $\langle A\rangle_l$ & $F_l$ & $F_l$ & $F_l$ & $A_{\rm max}$ & $A_{\rm max}$ & $A_{\rm max}$\\
\hline
$0\degr\leq i<15\degr $ & 0.15 & 0.23 & 0.41 & 0.11 & 0.13 & 0.17 & 0.95 & 0.87 & 0.68 & 1.51  & 1.61  & 1.63\\
$15\degr\leq i<30\degr$ & 0.17 & 0.25 & 0.44 & 0.12 & 0.13 & 0.16 & 0.93 & 0.85 & 0.65 & 1.66  & 1.68  & 1.71\\
$30\degr\leq i<45\degr$ & 0.19 & 0.28 & 0.49 & 0.12 & 0.13 & 0.14 & 0.91 & 0.82 & 0.61 & 1.93  & 2.04  & 1.98\\
$45\degr\leq i<60\degr$ & 0.25 & 0.36 & 0.61 & 0.13 & 0.13 & 0.13 & 0.86 & 0.77 & 0.55 & 2.56  & 2.56  & 2.60\\
$60\degr\leq i<75\degr$ & 0.41 & 0.54 & 0.87 & 0.15 & 0.14 & 0.11 & 0.77 & 0.69 & 0.51 & 4.22  & 4.28  & 4.34\\
$75\degr\leq i<90\degr$ & 1.24 & 1.40 & 1.78 & 0.16 & 0.14 & 0.08 & 0.56 & 0.54 & 0.50 & 27.79 & 27.34 & 20.51\\
\hline
\end{tabular}
\end{table*}

The main geometrical parameter governing the distribution of the total extinction $A_V$ is the inclination of the host galaxy on the sky.

This can be clearly seen from Figure \ref{avinc}, where the extinction
distribution is shown for six different galaxy inclination bins  for an S21 model with $\tau_V(0)=1.0$ and $B/T=0.3$. For the sake of clarity we reported in Table \ref{hatanostats} the mean extinction for the full sample $(\langle A\rangle_f)$ and for an extinction-limited subset of objects with $A_V<0.6$ mag $(\langle A\rangle_l)$. As expected, the effect of extinction in the face-on sample is modest though there is a significant probability, $P(A_V>0.6)\simeq 13\%$ (see Tab. \ref{hatanostats}), to have individual extinctions higher than 0.6 mag. In the edge-on sample the fraction of highly extinguished SNe is higher, as it can be seen from the extended high-extinction tail in the $A_V$ distribution. In this case it is $P(A_V>0.6)\simeq 46\%$. These two results are easily explained in terms of the maximum optical depth a Type Ia SN can experience because of the combination of its position inside of the galaxy and the line of sight along which the galaxy is seen by the observer. What is less intuitive is that in edge-on galaxies there is still an high probability that a Type Ia SN suffers only a small amount of extinction $P(A_V<0.1)\simeq23\%$.
The explanation for such effect is again found in geometrical arguments: the dust is confined on a disk which is thinner than the disk and the bulge (see Tab. \ref{modpars}) on which Type Ia SNe are distributed. Thus, as long as the SN height above (or below) the galaxy plane is larger than the dust scale-height, the Type Ia SN will suffer only little, and in some cases even negligible, absorption. If we consider the extinction-limited sample (see Tab. \ref{hatanostats}, col 6), we see that the mean extinction still increases with the inclination but only slightly: the net effect is only of a $\sim0.01$ mag increase from the face-on to the edge-on case.

As a final test we performed the same analysis described above for bulge and disk SNe separately. For this purpose we run another two simulations, one with $B/T=0.0$ to generate all the SNe in the disk and one with $B/T=1.0$ to generate all the SNe in the bulge. We choose this approach instead of analysing the two SN population of the $B/T=0.3$ simulation in order to keep the analysis at the same statistical significance level. The figures are listed in Tab. \ref{hatanostats}. The mean absorption for the full sample, $\langle A\rangle_f$, increases with increasing inclination angles, as for the $B/T=0.3$ case discussed above, but for disk SNe the mean value at each inclination is always smaller than for bulge SNe. The explanation for this apparently strange behaviour should again be found in geometrical arguments: the fact that the mean extinction value at any inclination is always larger for bulge SNe is due to the fact that those objects are located in the region where the dust density is higher. Indeed some of the disk SNe will be located at a large radial distance, where the dust density is very low, and thus they suffer large amounts of extinction only for a very small fraction of lines of sight. If we consider the extinction-limited subset we find that the mean extinction value increases mildly with the inclination showing a 0.06 mag enhancement from the face-on to the edge-on case. Conversely, for bulge SNe the trend is opposite: the mean extinction decreases with increasing inclinations. Once again the explanation is geometrical: the dust is confined in a disk which is much thinner then the bulge and thus with increasing inclination of the host galaxy the fraction of SN seen through the denser part of the disk is progressively reduced. Finally, we note that HBD98 model's predictions agree only partially with ours. In particular, they found that the mean extinction for the extinction limited sample decreases with increasing inclinations both for bulge and disk SNe. Given that their model predicts a decrease of only 0.04 mag from the face-on to the edge-on case for disk SNe, the difference is possibly due to the assumed central cut-off in their dust density distribution function (see also Section \ref{dust}).

\subsection{Extinction distribution as a function of the $B/T$ ratio}\label{sec:extdist_vs_b2t}

\begin{figure}
%
% The following figure was produced with:
% analsim.x -dextdist S21-3.1__t1.0__b0.0.sim
%		S21-3.1__t1.0__b0.5.sim -avmax 7. -avstep 0.1 -distype eps 
%		-paper -plotfile S21-3.1__t1.0__avdist__b2t
%
\includegraphics[width=84mm]{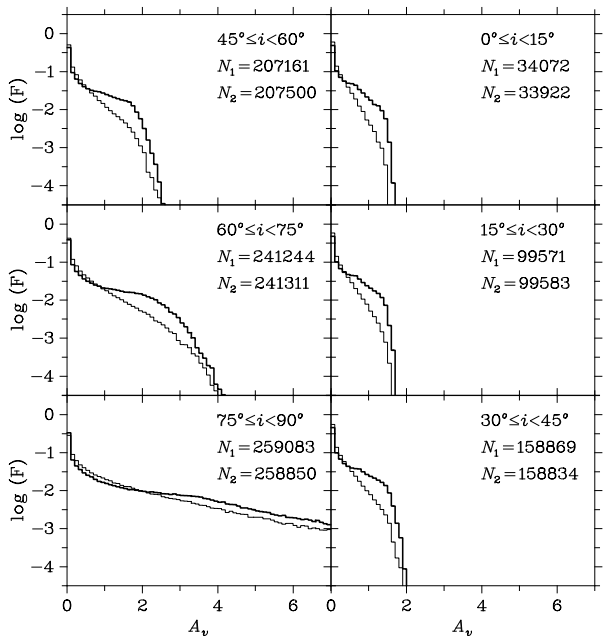}
\caption{Distribution of the total extinction $A_V$ for different galaxy inclination angles and the two extreme values of $B/T$: 0 (thin line) and 0.5 (thick line) for an S21 model with $\tau_V(0)=1.0$ and $B/T=0.3$. The fraction $F$ of objects in the inclination bin considered is plotted in log scale. In each panel the range of inclinations considered and the total number of simulated galaxies is shown. For the sake of clarity, the plot has been  intentionally limited to the range $0\leq A_V\leq7$ and to $10^{-5}\leq F\leq 1$. The number of SNe in each inclination bin is shown for both cases $N_1(B/T=0.0)$, $N_2(B/T=0.5)$.}
\label{avb2t}
\end{figure}

In Figure \ref{avb2t} we show the $A_V$ distributions for six different inclination bins and the two extreme values of $B/T$: 0.0 (thin line) and 0.5 (thick line) for an S21 model with $\tau_V(0)=1.0$ and $B/T=0.3$. Let us first consider, for each inclination bin, the peak of the corresponding distribution. With increasing inclination angles the relative fraction of SNe with $A_V<0.1$ mag (the bin's width) in a bulge-less system and in a $B/T=0.5$ system $(F_{0.0}/F_{0.5})$ is reverted passing from 1.24 in face-on galaxies $(0\degr\leq i<15\degr)$ to 0.87 in edge-on galaxies $(75\degr\leq i<90\degr)$.
This means that the presence of a bulge has a different effect in edge-on and face-on spirals. In the former case, the presence of the bulge increases the fraction of SNe with low extinction because the probability of a SN being at a larger distance from the galaxy plane is increased. On the contrary, in the face-on sample the presence of the bulge decreases the fraction of SNe with low extinction because the probability of a SN being above, or below, the disk (i.e. crossing a large section of the disk, where the dust is located) is increased. 

Another interesting feature that appears from Fig. \ref{avb2t} is that, regardless of the inclination bin considered, the $B/T=0.5$ model (thick line) appears to follow the $B/T=0.0$ one for a while and then becomes larger at a certain value of $A_V$ which depends on the inclination bin considered. This is clearly due to the fact that, in the $B/T=0.5$ model, a fraction of the SNe which explode in the bulge happen to be located also inside the disk, thus explaining the increased fraction of objects with higher extinction with respect to the bulge-less model.

\subsection{Projected radial distributions}

\begin{figure}
%
% The following figure was produced with:
% analsim.x -prjplot S21-3.1__t1.0__b2t0.3__z0.15__1E6.sim
%		-paper -distype eps -plotfile S21-3.1__t1.0__b0.3__prjdist
%
\includegraphics[width=84mm]{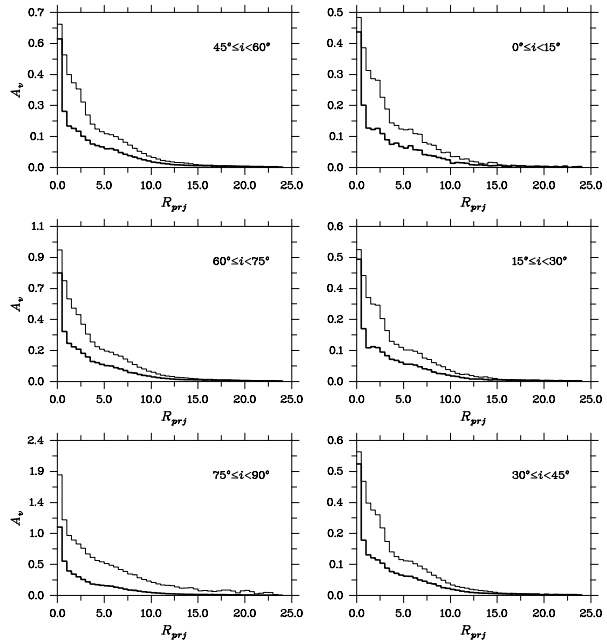}
\caption{Mean (thin line) and median (thick line) $A_V$ as a function of the projected radial distance $R_{prj}$ from the centre of the host galaxy for an S21 model with $\tau_V(0)=1.0$ and $R_V=3.1$.}
\label{avproj}
\end{figure}

An interesting quantity to analyse is the projected radial behaviour of the mean and median absorption. The interest clearly arises from the possibility of comparing the predicted distributions with the observed ones.
In Figure \ref{avproj} we show the mean (thin line) and median (thick line) $A_V$ projected radial distributions for a S21 model with $\tau_V(0)=1$ and $B/T=0.3$. The two distributions have been constructed by selecting all SNe located inside concentric circular bins, at increasing projected radial distance, and then computing, for each subset, the mean and the median extinction values. As expected, the mean and median absorption values in the nuclear region increase with larger inclination angles of the host galaxy. If we consider the distribution of median absorption, we see that the radial functional dependency is quite similar for each inclination, the only feature changing being the peak value of the distribution. We note also that, in each inclination bin considered, SNe suffering high extinction are always those projected in the very central regions. Finally, the fact that the median distribution is smaller than the mean one suggests that, at a given radial distance, there is a "tail" of SNe suffering high extinctions (as implied by the definition of mean and median).

\subsection{The spiral perturbation of the dust disk}

\begin{table}
\caption{The effect of dust spiral geometry on SN magnitudes. The first column indicates the two models for which the difference $|\Delta m_V|$ was considered. The next columns report the fraction of objects for which the magnitude difference obtained in the two models is lower than 0.025 mag, or greater than 0.2 mag and 1.0 mag.}
\label{tab:spiral}
\begin{tabular}{lccc}
Geometries & $|\Delta|<0.025$ & $|\Delta|\geq0.2$ & $|\Delta|\geq 1.0$\\
\hline
(S0-S21)  & 0.50 & 0.14 & 0.003\\
(S0-S22)  & 0.44 & 0.22 & 0.009\\
(S0-S31)  & 0.52 & 0.11 & 0.002\\
(S0-S32)  & 0.45 & 0.19 & 0.006\\
(S21-S31) & 0.51 & 0.17 & 0.007\\
(S22-S32) & 0.45 & 0.23 & 0.007\\
\hline
\end{tabular}
\end{table}

\begin{figure}
%
% The following figure was produced with:
% analsim.x -geo S0-3.1__t1.0__b0.3__z0.10__1E6.sim 
%		S21-3.1__t1.0__b0.3__z0.10__1E6.sim
%		S22-3.1__t1.0__b0.3__z0.10__1E6.sim
%		S31-3.1__t1.0__b0.3__z0.10__1E6.sim
%		S32-3.1__t1.0__b0.3__z0.10__1E6.sim
%		-histmax 3.025 -plotmax 0.75 -hstep 0.05 -paper
%		-distype eps -plotfile spiral__geometry__t1.0
%
\includegraphics[width=84mm]{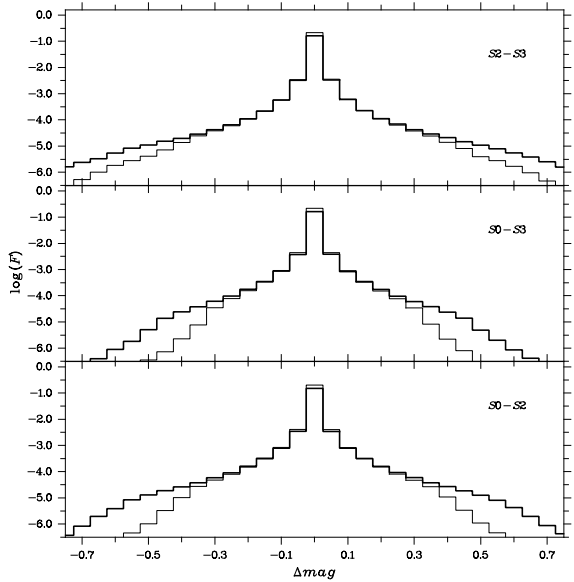}
\caption{Effect of different dust geometries on simulated SN magnitudes. The three panels show the distribution of the difference between the observed magnitude, for the same SN positions and line of sights, in the S0, S2 and S3 models. The fraction $F$ of objects is plotted in a log. scale. The thin lines show the distributions of $\Delta m_V(S0-S21)$ (bottom panel), $\Delta m_V(S0-S31)$ (middle panel) and $\Delta m_V(S21-S31)$ (top panel) whilst the thick lines show the distributions of $\Delta m_V(S0-S22)$ (bottom panel), $\Delta m_V(S0-S32)$ (middle panel) and $\Delta m_V(S22-S32)$ (top panel). All distributions have a bin size of $0.05$ mag.}
\label{dustgeo}
\end{figure}

We performed a simple test to quantify the impact the choice of one dust geometry with respect to another has on the observed SN magnitudes. We run one Monte Carlo simulation of $10^6$ SN positions and line of sights for a galaxy with $B/T=0.3$ (the only model free parameter affecting the distribution of Type Ia SNe). We then computed the observed SN magnitudes for each of the five dust geometries summarised in Table \ref{dustpars} adopting $\tau_V(0)=1.0$. In this way, SN positions are the same in each simulation and thus their magnitudes obtained in each dust geometry can be directly compared.
In particular, we show in Figure \ref{dustgeo} the magnitude difference $\Delta m_V$ between: the plain disk dust distribution S0 and each of the two spiral arms ones (S21, S22; bottom panel); the plain disk dust distribution S0 and each of the three spiral arms one (S31, S32; middle panel); each of the two spiral arms dust distributions and the corresponding three spiral arms ones (upper panel). From this test we can conclude that:
\begin{itemize}
\item
in each of the four cases shown in the bottom and middle panels of Fig. \ref{dustgeo}, the effect introduced by the spiral perturbation of the dust disk with respect to the plain disk case on the observed magnitudes is negligible for a fraction of SNe which goes from a minimum of $\sim44\%$ (when the S22 geometry is considered), to a maximum of $\sim52\%$ (when the S31 geometry is considered), being $|\Delta m_V| < 0.025$ mag.

\item
the fraction of SNe with $|\Delta m_V|\geq0.2$ mag, i.e. those showing a significant magnitude difference, is larger for the extreme spiral cases (S22, S32) than for the "normal" spiral cases (S21, S31). Moreover the difference between the plain dust disk and the spiral perturbed one is always larger in the two-arms cases with respect to the corresponding three-arms ones. The figures are shown in Table \ref{tab:spiral}. Finally, the fraction of SNe with $|\Delta m_V|>1.0$ mag is lower then 1\% in all the cases.
\end{itemize}

\section{Type Ia SN rate correction factor}\label{par:app}

To exemplify how our model can be used to derive the extinction correction factor, we will analyse a single and simple test case. In particular, we will consider an ideal SN search whose galaxy sample is made of objects of the same type, at the same distance and with the same properties ($R_V$, $\tau_V(0)$, $B/T$ and dust geometry). 
This is almost equivalent to consider a real SN search targeted on a galaxy cluster for which the calculation of the correction factor is restricted to a single galaxy morphological class. It seems indeed plausible that for next future SN searches the statistics will no longer be a limitation and hence the computation of the SN rates in different galaxy types will become feasible. We thus run a simulation of $10^7$ Type Ia SNe in a galaxy with $\tau_V(0)=1.0$, $B/T=0.3$ at redshift $z=0.10$ and with the dust geometry we labelled S21 (see Tab. \ref{modpars}).

It is obvious that the effect of host extinction is to reduce the observed SN number counts and thus the derived SN rates will be only a lower limit if no correction is applied. To compute the SN rate extinction correction factor $\mathcal{R}$ from our simulation data we just need the intrinsic magnitude distribution $N_0(m)$ and the extinguished magnitude distribution $N_e(m)$. Then, $\mathcal{R}$ will be simply given by the following relation:
\begin{equation}\label{eq:corrfac}
\mathcal{R} = \frac{\displaystyle\int_{m_1}^{m_{\epsilon}}{N_0(m)\epsilon(m)dm}}
{\displaystyle\int_{m_1}^{m_{\epsilon}}{N_e(m)\epsilon(m)dm}}
\end{equation}
where $\epsilon(m)$ is the SN search detection efficiency function (see below) and the integrals are extended over the magnitude range defined by the bright end of the magnitude distribution $m_1$ and the faintest magnitude for which the efficiency function is not null $m_{\epsilon}$. Extinction corrected SN rates are eventually obtained simply multiplying the uncorrected rates by the correction factor $\mathcal{R}$.
The SN detection efficiency function, $\epsilon(m)$, provides the sensitivity of the SN search as a function of the SN candidate's magnitude. The detection efficiency as a function of SN magnitude is usually a complicated function of many parameters, like the quality of the subtracted images (seeing, transmission) and the technique (convolution, selection criteria) used to extract the SN candidates. The efficiency function is usually determined through Monte Carlo simulations in which a number of artificial SNe are positioned on the search images \citep[see][]{marco,stat04}. These synthetic images are then processed with the same software used to perform the search on the real images and the fraction of recovered--over--injected artificial SNe, $\epsilon=N_r/N_{\rm tot}$, is determined. In this work we used one of the efficiency curves of \citet{stat04} for a seeing of $0\farcs90$ re-scaled to an exposure time of 600 sec, which is a reasonable value for a $z=0.10$ SN search performed with a 2m-class telescope \citep[as in the case of][]{stat04}. We note that the effect of exposure time on the efficiency curve is equivalent to a rigid shift of the magnitude scale.

\begin{figure}
%
% The following figure was produced using the SN macro
% "paperfig" included in the file "paper.mac". This macro
% requires the input files: "mag_distribution.dat" and 
% "magabs_distribution.dat" which are produced by the
% command:
%	 analsim.x -SNmag S21-3.1__t1.0__b0.3__z0.10__1E7.sim
%
\includegraphics[width=84mm]{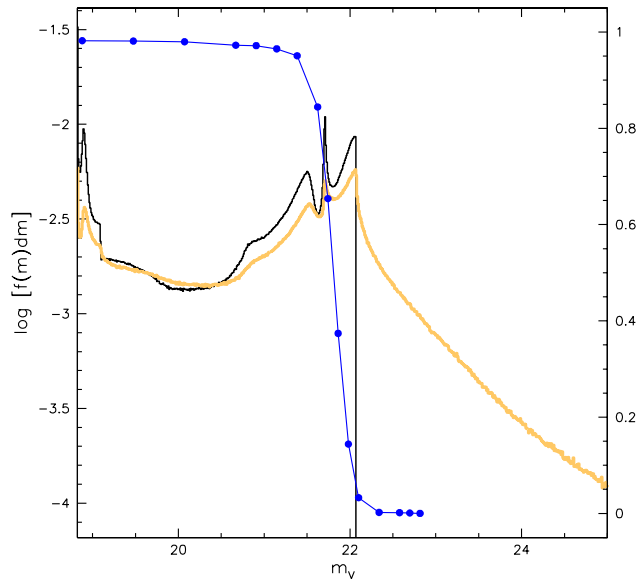}
\caption{Intrinsic, $f_0(m)dm$ (black line), and extinguished, $f_e(m)dm$ (grey line), magnitude distributions. The connected solid dots represent the SN detection efficiency curve used in this work (see the text for further details).}
\label{magdist}
\end{figure}

As an example, we have applied the method described at the beginning of the section. First we show in Fig. \ref{magdist} the intrinsic, $f_0(m)dm$ (black line), and the extinguished, $f_e(m)dm$ (grey line) magnitude distributions normalised to the total number of simulated objects. The shape of $f_0(m)dm$ is linked to that of the SN LC: at phases $\phi>30-40{\rm d}$ the LC is much less steep than at earlier phases and thus, for a given magnitude bin, there is a wide range in phase (i.e a large number of SNe) contributing to that bin. The spike and wiggles at $21\leq m_V\leq22$ are mainly due to the fact that in this simulation we used only one specific Type Ia LC. We expect them to smooth out when Type Ia SN diversity is taken into account by combining simulation with different SN properties.
From our calculation we find $\mathcal{R}=1.27$, i.e. $\sim27\%$ of SNe are not detected by the search because of host galaxy extinction. Given the impressive size of the tail of the $f_e(m)dm$ in Fig. \ref{magdist}, it is interesting to look at the intrinsic fraction (without taking into account the efficiency curve) of SNe that are extinguished beyond the fainter magnitude reached in the absence of extinction ($m_V\simeq22.08$ in the case shown in Fig. \ref{magdist}): the tail includes $\sim24\%$ of the simulated SNe.

\section{Discussion}\label{par:discussion}

The recipe described above to derive the correction factor $\mathcal{R}$ clearly depends on the model parameters. To assess their importance on the final result we performed several simulations to probe its dependence from: (1) dust geometry, (2) size of the bulge, (3) dust content and (4) dust type.

1. We computed the correction factor for each dust geometry described in Tab. \ref{dustpars} using $\tau_V(0)=1.0$, $R_V=3.1$ and $B/T=0.3$. We found that $\mathcal{R}=1.27$ in all cases. This result demonstrates that the distribution of dust does not play a major role and thus it is possible to adopt the simplest distribution, i.e. the plain double exponential disk S0. Though this is true when the extinction correction $\mathcal{R}$ is concerned, we note that a detailed dust distribution should be included when studying the distribution of extinction and its effect on SN properties/distributions (see Sec. \ref{par:parspace}). This result is not unexpected because (a) the spiral perturbation of the dust distribution creates regions of enhanced (the spiral arms) and lower density with respect to the plain exponential disk; and (b) Type Ia SNe are distributed (in the simulations) on a plain exponential disk, i.e. without a spiral perturbation. The effect of dust over- and under-densities therefore cancels out. Conversely we expect that the dust distribution will play a more significant role for CC SNe because they are also more concentrated in the spiral arms.

2. We computed the correction factor for $0.0\leq B/T\leq 0.5$ using dust model S21 with $\tau_V(0)=1.0$ and $R_V=3.1$. We found that the corresponding range spanned by the extinction correction factor is $1.22\leq\mathcal{R}\leq1.31$. Hence, the fraction of Type Ia SNe that are located in the bulge, which is governed by the $B/T$ parameter,  plays a role not only on the extinction distribution (see Sec. \ref{sec:extdist_vs_b2t}) but also on the extinction correction factor $\mathcal{R}$.

3. We computed the correction factor $\mathcal{R}(\tau_V)$ for different values of the total face-on optical depth $\tau_V(0)$ using dust model S0 with $R_V=3.1$ and $B/T=0.3$. We found: $\mathcal{R}(0.5)=1.16$, $\mathcal{R}(1)=1.27$, $\mathcal{R}(2.5)=1.57$, $\mathcal{R}(5)=1.91$ and $\mathcal{R}(10)=2.35$.

4. We computed the correction factor for $R_V=3.1$, 4.0 and 5.5 using dust model S0 with $\tau_V(0)=1.0$ and $B/T=0.3$. We recall that these are the values for which the extinction coefficients $C_{\rm ext}^H(\lambda)$ are available (see Sec. \ref{drainecext}). We found that the corresponding values of the extinction correction factor are $\mathcal{R}=1.27$, 1.31, 1.34.

When the determination of $\mathcal{R}$ is concerned, although the detailed geometry of dust can be neglected, the dust content, properties and the bulge size still play a major role. The computation of $\mathcal{R}$ for a real SN search requires indeed several simulations: one for each "galaxy class" which is defined by the tuple $(B/T, R_V, \tau_V)$. If the SN rates cannot be computed (and hence corrected) for each galaxy class separately, a global correction factor $\langle\mathcal{R}\rangle$ should be derived from a weighted mean of the $\mathcal{R}_i$ obtained for each galaxy class $i$, with weights given by the fraction of objects in each galaxy class.

\section{Summary}\label{par:end}

We presented a new Monte Carlo approach to the problem of determining the effect of host galaxy extinction on observed SN properties and to compute the extinction correction factor for SN rate. Although we provided a detailed discussion of the role played by the model free-parameters, our main purpose was to outline the recipe to derive the extinction correction factor $\mathcal{R}$ for Type Ia SN. Our main results are that (1) the value of $\mathcal{R}$ is almost insensitive to the particular spiral geometry of dust and hence a plain exponential disk is a good approximation. (2) the value of $\mathcal{R}$ strongly depends on the galaxy physical properties, in particular the total face-on optical depth $\tau_V(0)$, the value of $R_V$ and the size of the bulge $B/T$. If we adopt a Milky Way-like dust $(R_V=3.1)$, $B/T=0.3$ and the range of most likely total face-on optical depths $1\leq\tau_V(0)\leq5$ \citep{bianchirev}, the corresponding range spanned by the extinction correction factor is $1.27\leq\mathcal{R}\leq1.91$.

The computation of $\mathcal{R}$ for a real SN search involves a number of complications due to the heterogeneity of the galaxy sample, the redshift dependence of galaxy properties and finally the variety of Type Ia SNe. In particular, this work did not take into account neither the intrinsic spread in Type Ia SNe absolute magnitudes nor the correlation between the light curve shape (characterised, e.g. by the $\Delta m_{15}$ parameter) and the morphological types of SN hosts. Such simplifications were possible only because the aim of our work was to provide an overview of the method rather than discussing the application to a real case, which will be instead the subject of a future work.
This application clearly requires a detailed study of the observed galaxy sample and the definition of the corresponding model free-parameters which was beyond the objective of the present study. Finally, we note that, though the approach described in this work can be used also for core collapse SNe, its feasibility is limited both by the lack of adequate spectral libraries and by the extreme heterogeneity of these objects.

\section*{Acknowledgments}
We thank the two referees, Sidney van den Bergh and David Branch for their useful comments which improved the quality of this paper.
We would like to thank also Enrico Cappellaro, Bruno Leibundgut and Brian Schmidt for reading the manuscript before the submission and for their helpful suggestions and ideas. MR would like to thank Giuliano Pignata for helpful discussions on synthetic photometry. MR was supported by an ESO Studentship and acknowledges the generous hospitality of ESO headquarters, where this paper was conceived.
%
%%%%%%%%%%%%%%%%%%%%%%%%%%%%%%%%%%%%%%%%
%

\label{lastpage}

\end{document}